\documentclass{cernrep}

\begin{document}

\title{Photon-photon and photon-hadron processes in \textsc{Pythia}~8}

\author{I. Helenius}
\institute{Institute for Theoretical Physics, T\"{u}bingen University, Auf der Morgenstelle 14, 72076 T\"{u}bingen, Germany}

\begin{abstract}
We present a new implementation of photoproduction processes in e$^+$e$^-$ and ep collisions into \textsc{Pythia}~8 Monte-Carlo event-generator. In particular we discuss how the parton showers and multiparton interactions are generated with a resolved photon beam and what is the relative contribution from direct processes in different kinematical regions. As an application we show compa\-risons to data for charged-particle production in e$^+$e$^-$ and ep collisions at LEP and HERA. We consider also photoproduction of dijets comparing to data for ep collisions at HERA and discuss about possibility to further constrain nuclear PDFs with ultra-peripheral heavy-ion collisions at the LHC.
\end{abstract}


\maketitle

\section{Introduction}

Photon-initiated processes can be studied in many different collider setups. In the future e$^+$e$^-$ colliders photon-photon processes can generate additional QCD background for  many processes and, for example, provide an additional channel to produce a Higgs boson. In ep colliders one can study photon-hadron interactions which are sensitive to the structure of the resolved photon and the target hadron, and contribution of the multiparton interactions (MPIs) for particle production with photon beams. Furthermore, different nuclear modifications can be probed in ultra-peripheral heavy-ion collisions. Here there are no hadronic interactions but one of the nuclei emit a photon that interact with the other nucleus. In particular these collisions can provide important future constraints for nuclear PDFs (nPDFs).

\textsc{Pythia}~8 \cite{Sjostrand:2014zea} is a general purpose Monte-Carlo event generator that is capable of simulating all particles created in an event. Event generation starts from the hard process of interest. The next step is to generate initial- (ISR) and final-state radiation (FSR) and the MPIs, evolving from the hard-process scale down to a scale below which physics becomes non-perturbative, see \Ref~\cite{Sjostrand:2004ef} for details. The event is then hadronized using Lund string model and unstable hadrons are decayed into stable ones measured in the detector. Main emphasis has been on pp collisions at the LHC but extensions to other collision systems have been developed. Here we discuss about recent developments for photon-photon and photon-hadron interactions in e$^+$e$^-$ and ep collisions \cite{HeleniusSjostrand}.

\section{Framework}

Probability for the MPIs in \textsc{Pythia}~8 is given by the $2\rightarrow 2$ QCD processes \cite{Sjostrand:2004pf}. The divergence in the $p_{\mathrm{T}}\rightarrow 0\UGeVc$ limit is regulated with a screening parameter $p_{\mathrm{T0}}$ such that
\begin{equation}
\frac{\mathrm{d}\sigma}{\mathrm{d}p_{\mathrm{T}}^2} \propto \frac{\alpha_{\mathrm{S}}(p_{\mathrm{T}}^2)}{p_{\mathrm{T}}^4} \rightarrow \frac{\alpha_{\mathrm{S}}(p_{\mathrm{T}}^2+p_{\mathrm{T0}}^2)}{(p_{\mathrm{T}}^2 + p_{\mathrm{T0}}^2)^2}.
\end{equation}
The parameter is taken to be energy dependent and is parameterized as $p_{\mathrm{T0}}(\sqrt{s}) = p_{\mathrm{T0}}^{\mathrm{ref}} (\sqrt{s}/7\UTeV)^{\alpha}$, where the default Monash-tune provides values $p_{\mathrm{T0}}^{\mathrm{ref}}=2.28\UGeVc$ and $\alpha = 0.215$ for (anti)proton beams. Since the structure of a resolved photon is evidently different than the structure of a proton, the value of the screening parameter should be revised. This is one of the outcomes of the presented work.

\subsection{Photon beam}

Photons may interact as an unresolved particle (direct photon), or fluctuate into a hadronic state with equal quantum numbers (resolved photon). In the former case the photon itself act as an initiator of the hard process whereas in the latter case the constituent partons are the initiators. As with hadrons, the distribution of the partons can be described with PDFs, $f^{\gamma}_i(x,Q^2)$, which scale evolution are given by the DGLAP equations. In addition to the usual splittings of partons, for a resolved photon one needs to take into account also $\gamma \rightarrow q \bar{q}$ splittings of the beam photon, giving \cite{DeWitt:1978wn}
\begin{equation}
\frac{\mathrm{\partial} f^{\gamma}_i(x,Q^2)}{\mathrm{\partial}\,\mathrm{log}(Q^2)} = \frac{\alpha_{\mathrm{EM}}}{2\pi}e_i^2 P_{i\gamma}(x) + \frac{\alpha_{\mathrm{S}}(Q^2)}{2\pi} \sum_j \int_x^1\frac{\mathrm{d}z}{z}\, P_{ij}(z)\, f^{\gamma}_j(x/z,Q^2),
\label{eq:gammaDGLAP}
\end{equation}
where $P_{ij}(z)$'s are the usual DGLAP splitting kernels for a given $j\rightarrow i k$ splittings. The additional $\gamma \rightarrow q \bar{q}$ splittings provide more quarks at higher values of $x$ than with hadron beams. In this work we use photon PDFs from CJKL analysis \cite{Cornet:2002iy}. For the parton shower generation the additional term corresponds to a probability to end up to the original beam photon when tracing back the ISR splittings that have taken place for the hard-process initiators. If this happens, there are no further ISR emissions or MPIs below the scale where this happens, or need for any beam remnants.

Since the interacting photons can be either resolved or direct, there are three different types of processes that needs to be taken into account for a photon-photon interaction: resolved-resolved, resolved-direct, and direct-direct. For the resolved-resolved contribution the full parton-level evolution needs to be generated including ISR and FSR and also possible MPIs. For direct-resolved case no MPIs are present nor ISR for the direct side. For direct-direct case only FSR is relevant. The relative contribution of each process type depends on the kinematics, typically direct (resolved) processes dominate when $x$ is large (small). In case of photon-hadron interaction the photons can be either resolved or direct.

\subsection{Photon flux from leptons}

The photon flux from lepton $l$ can be modelled with equivalent photon approximation 
(EPA). Integrating the flux over allowed photon virtuality yields
\begin{equation}
f_{\gamma}^l(x_{\gamma},Q_{\mathrm{max}}^2) = \frac{\alpha_{\mathrm{EM}}}{2\pi} \frac{1+(1-x_{\gamma})^2}{x_{\gamma}} \log \left( \frac{Q_{\mathrm{max}}^2}{Q^2_{\mathrm{min}}(x_{\gamma})} \right).
\label{eq:EPA}
\end{equation}
The virtuality of the photon, $Q^2$, is related to the lepton scattering angle so the lower $Q^2$ limit for the splitting can be derived from the kinematics. The appropriate upper limit depends experimental configuration. Here we have considered only quasi-real photons so $Q_{\max}^2 \lesssim 1\UGeV^2$.

For the direct contribution the spectrum of photons can be obtained directly from the flux. The distribution of partons in resolved photons that is coming from the lepton beam can be obtained by convoluting the photon flux $f_{\gamma}^l$ with the parton-inside-photon PDFs $f_{i}^\gamma$
\begin{equation}
xf_{i}^l(x,Q^2) = \int_x^1 \frac{\mathrm{d}x_{\gamma}}{x_{\gamma}} x_{\gamma} f^l_{\gamma}(x_{\gamma},Q_{\mathrm{max}}^2) x'f^{\gamma}_i(x',Q^2),
\end{equation}
where $Q^2$ is now the factorisation scale and $x'=x/x_{\gamma}$. The partonic evolution is then performed for the photon-photon(hadron) sub-collision constructed according to sampled $x_{\gamma}$ values.
\section{Results}

\subsection{Charged-hadron photoproduction in e$^+$e$^-$}

Photon-photon interactions in e$^+$e$^-$ collider have been studied in LEP. A suitable observable to compare our new framework is the charged hadron production for which measurement from OPAL experiment exists \cite{Ackerstaff:1998ib}. The measurement used anti-tagged events where the angle of scattered leptons is beyond the acceptance so that they are not seen in the detector. With the OPAL kinematics this translates into a virtuality cut $Q^2 < 1\UGeV^2$.

\begin{figure}[ht]
\begin{center}
\includegraphics[width=\textwidth]{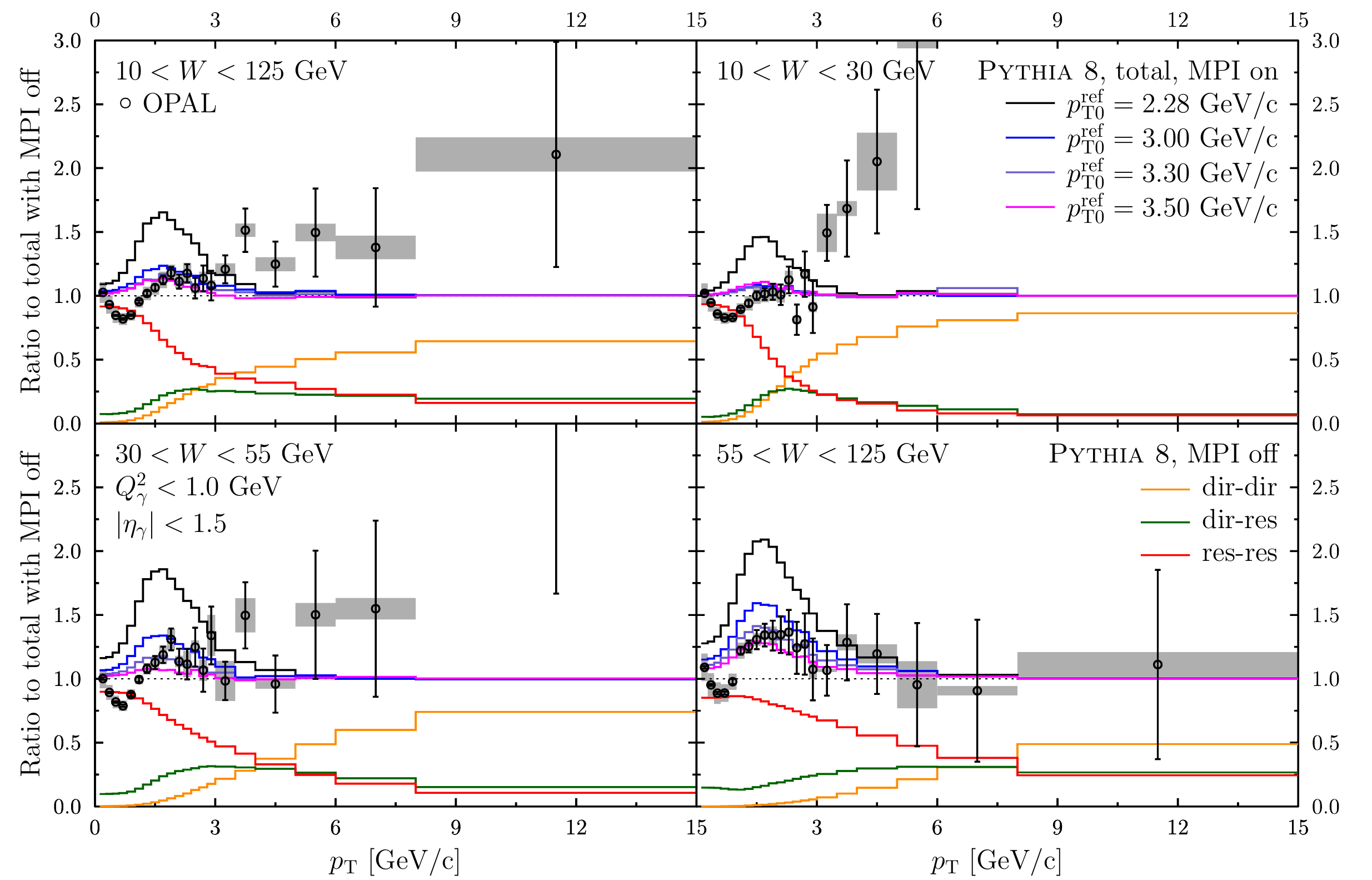}
\caption{Ratio between measured \cite{Ackerstaff:1998ib} and simulated charged-hadron cross sections in $\gamma\gamma$ interactions in e$^+$e$^-$ collisions at $\sqrt{s}=166\UGeV$ for different invariant mass bins, $10<W<125 \UGeV$ (top left), $10<W<30\UGeV$ (top right), $30<W<55\UGeV$ (bottom left) and $55<W<125 \UGeV$ (bottom right). Contribution from direct-direct (orange), direct-resolved (green) and resolved-resolved without MPIs (red) are shown separately and the sum of these are shown for different values of $p_{\mathrm{T0}}^{\mathrm{ref}}$.}
\label{fig:eeCh}
\end{center}
\end{figure}

\Figure[b]~\ref{fig:eeCh} shows the ratio between the OPAL data and the result of simulations combining direct and resolved contributions without any MPIs with different invariant mass $W$ bins. Also the individual contributions from direct-direct, direct-resolved and resolved-resolved are plotted separately to quantify the contribution from each of these. The data is then compared to the results with MPIs using different values for the parameter $p_{\mathrm{T0}}^{\mathrm{ref}}$. The comparison shows that the default $p_{\mathrm{T0}}^{\mathrm{ref}}=2.28\UGeVc$ generates too many charged particles around $p_{\mathrm{T}}\sim 2\UGeVc$. Increasing the value of this parameter reduces the number of charged particles in this region and a good agreement in all $W$ bins is obtained with $p_{\mathrm{T0}}^{\mathrm{ref}}=3.30~\UGeVc$. The systematic increase of the number of charged particles from the MPIs with increasing $W$ is supported by the data.

\subsection{Charged-hadron and dijet photoproduction in ep}

\begin{figure}[ht]
\begin{center}
\includegraphics[width=0.49\textwidth]{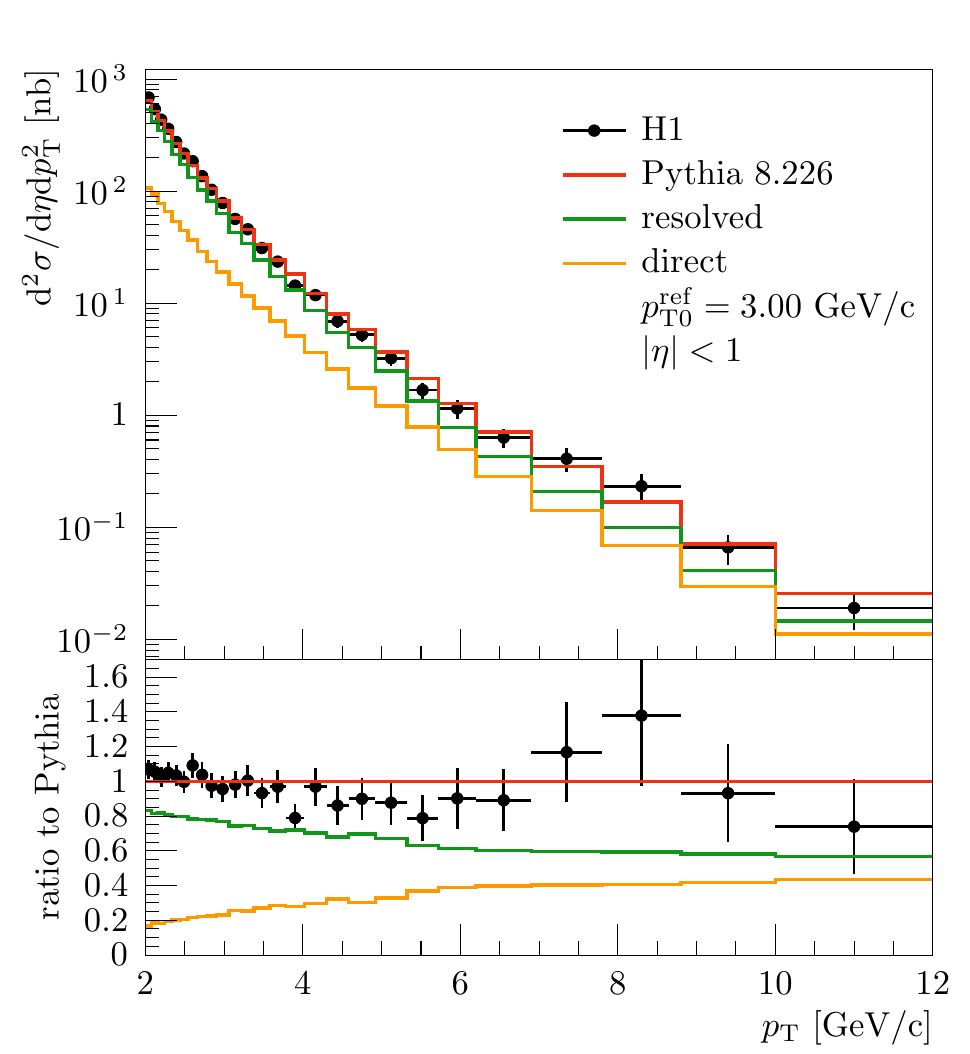}
\includegraphics[width=0.49\textwidth]{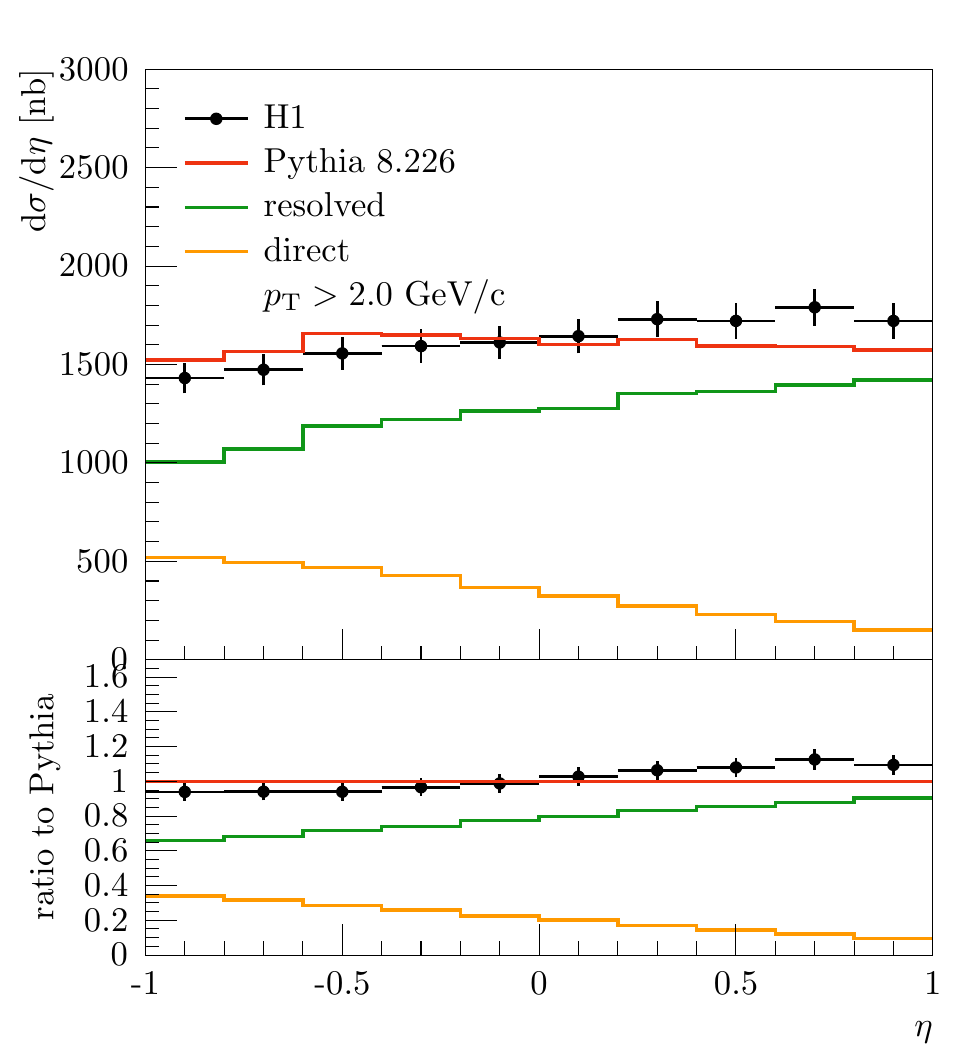}
\caption{Cross section for charged particle photoproduction as a function of $p_{\mathrm{T}}$ (left panel) and $\eta$ (right panel) in ep collisions at HERA at mid-rapidity. Data from H1  \cite{Adloff:1998vt} are compared to \textsc{Pythia} simulations with $p_{\mathrm{T0}}^{\mathrm{ref}} = 3.0 \UGeVc$ (red), decomposed to direct (orange) and resolved (green) contributions.}
\label{fig:ep}
\end{center}
\end{figure}

There are plenty of data available for the photoproduction in ep collisions from HERA collider. Again a useful observable to study the effect from MPIs is the charged-hadron production for which data exists from H1 \cite{Adloff:1998vt} and ZEUS \cite{Derrick:1995jq} experiments. The kinematical cuts in the H1 measurements corresponds to average invariant mass of photon-proton system of $\langle W_{\mathrm{\gamma p}}\rangle  = 200 \UGeV$. The simulations are compared to this data in \Fref{fig:ep} as a function of $p_{\mathrm{T}}$ and $\eta$. The data is best described with $p_{\mathrm{T0}}^{\mathrm{ref}} = 3.0\UGeVc$ which conveniently lie between the values optimal for $\gamma\gamma$ and pp. This difference in $p_{\mathrm{T0}}^{\mathrm{ref}}$ values could reflect that the photon is a cleaner state than the proton, but also that a more sophisticated energy scaling of $p_{\mathrm{T0}}(\sqrt{s})$ may be required, also for protons.

\begin{figure}[ht]
\begin{center}
\includegraphics[width=0.49\textwidth]{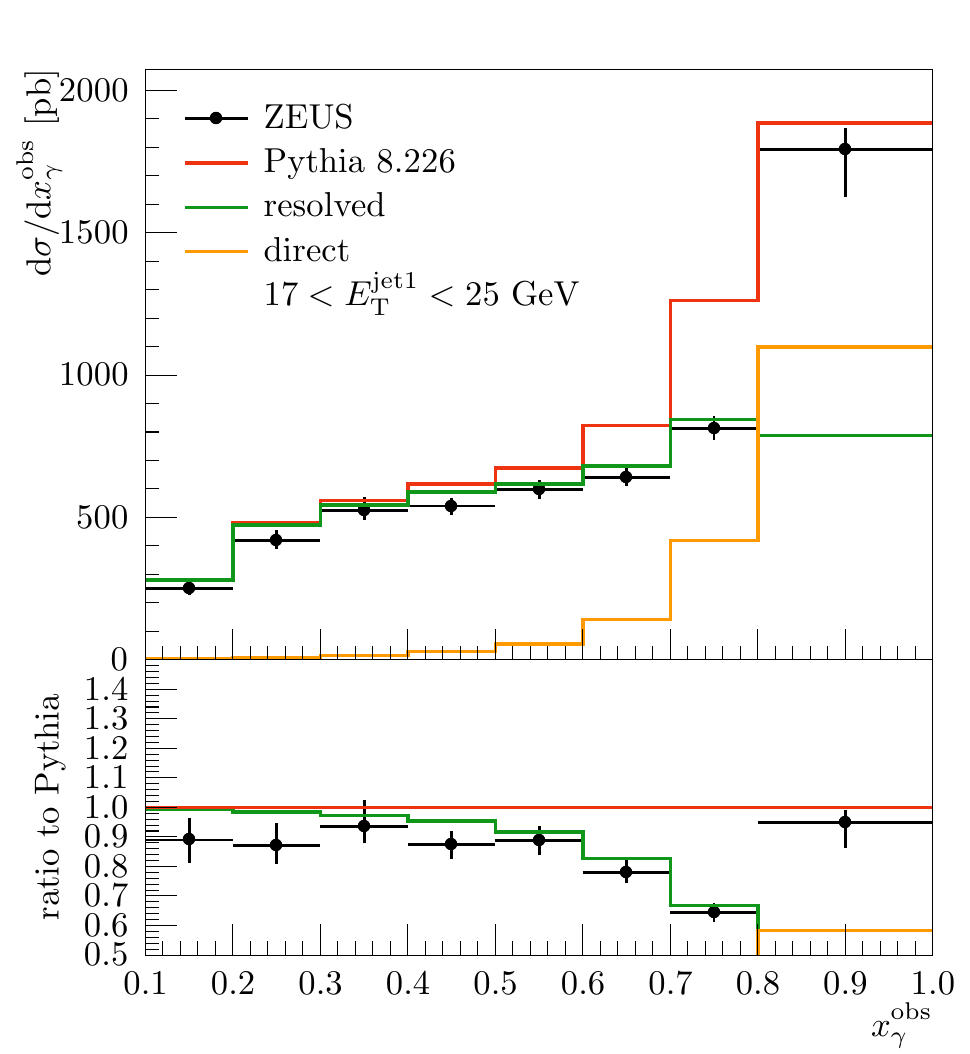}
\includegraphics[width=0.49\textwidth]{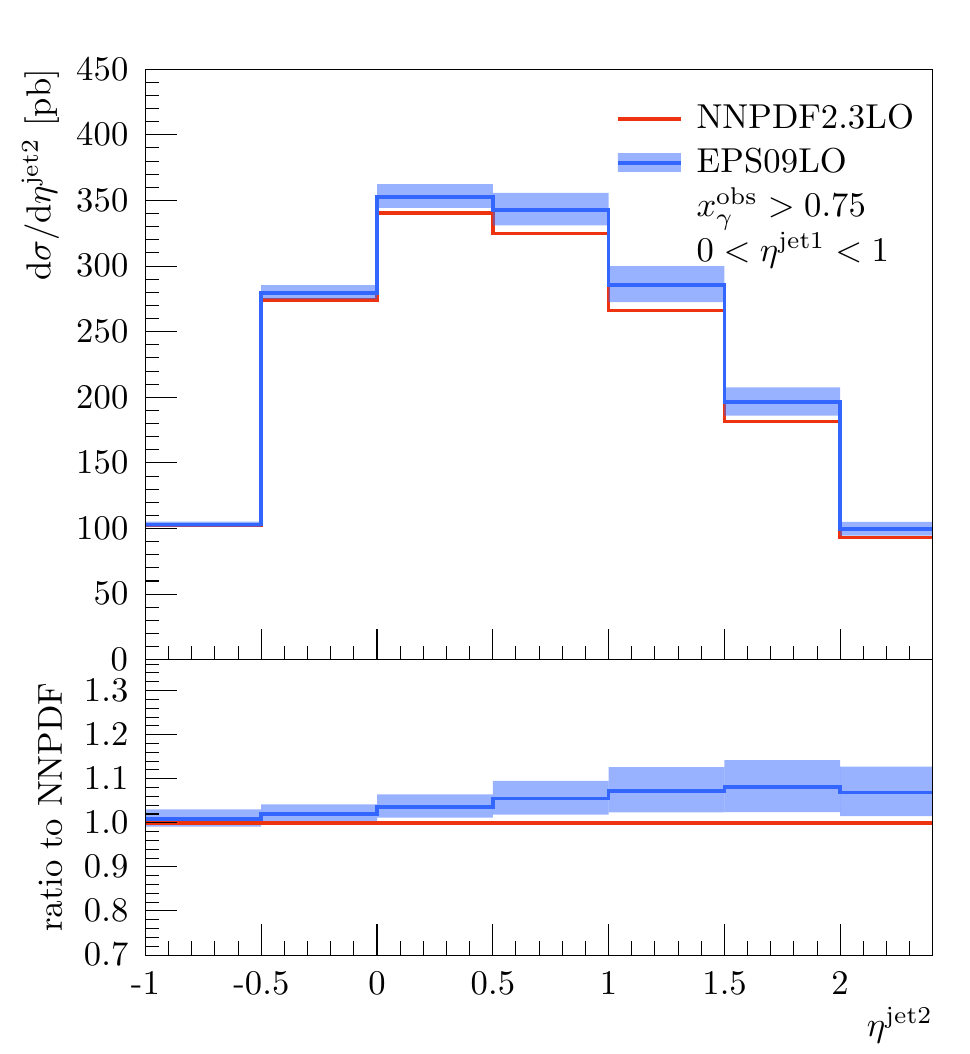}
\caption{\textbf{Left:} Cross section for dijet photoproduction in HERA as a function of $x_{\gamma}^{\mathrm{obs}}$. Data from ZEUS \cite{Chekanov:2001bw} are compared to \textsc{Pythia} simulations (red), decomposed to direct (orange) and resolved (green) contributions. \textbf{Right:} Cross section for dijet photoproduction as a function of $\eta^{\mathrm{jet2}}$ for events with $x_{\gamma}^{\mathrm{obs}}> 0.75$ using a proton PDFs only (red) and with nuclear modifications from EPS09 (blue) including the nPDF uncertainties (blue band).}
\label{fig:epDijet}
\end{center}
\end{figure}

Another useful observable to study photon-hadron interactions is photoproduction of dijets. For this we can use data from ZEUS \cite{Chekanov:2001bw} where the photon virtuality is restricted to $Q^2<1.0\UGeV^2$ and $134 < W_{\gamma \mathrm{p}} < 277\UGeV$. The transverse energy cuts in the data are $E_{\mathrm{T}}^{\mathrm{jet1}} > 14\UGeV$ and $E_{\mathrm{T}}^{\mathrm{jet2}} > 11\UGeV$, where jet 1 (2) is chosen to be the jet with the (second) highest $E_{\mathrm{T}}$ within pseudorapidities $-1<\eta<2.4$. Still it is not possible to separate the direct and resolved processes apart, but by defining
\begin{equation}
x_{\gamma}^{\mathrm{obs}} = \frac{E_{\mathrm{T}}^{\mathrm{jet1}}\mathrm{e}^{\eta^{\mathrm{jet1}}} + E_{\mathrm{T}}^{\mathrm{jet2}}\mathrm{e}^{\eta^{\mathrm{jet2}}}}{2yE_{\mathrm{e}}}
\end{equation}
some sensitivity for different contributions can be obtained. Here $y$ is the inelasticity of the event and $E_\mathrm{e}$ is the energy of the positron beam. \Figure[b]~\ref{fig:epDijet} shows a comparison of data and \textsc{Pythia} simulations for the dijet cross section as a function of $x_{\gamma}^{\mathrm{obs}}$, where again the direct and resolved contributions are shown separately. The simulations are performed with $p_{\mathrm{T0}}^{\mathrm{ref}} = 3.0\UGeVc$ tuned to charged particle production data above. In general the agreement is decent and indeed the events from direct processes tend to sit at higher values of $x_{\gamma}^{\mathrm{obs}}$ as expected. However, the resolved processes do provide some contribution also at $x_{\gamma}^{\mathrm{obs}}>0.8$. A possible explanation of the slight overshoot of the data might result from differences in the applied jet algorithms---this will be studied in more detail later on.

Recently it has been argued that dijet production in ultra-peripheral heavy-ion collisions at the LHC could provide further constraints for nuclear modification of the PDFs \cite{ATLAS:2017kwa}. Since the expected (per-nucleon) invariant mass of photon-ion system is not that far from the $W_{\gamma \mathrm{p}}$ in ep at HERA, a qualitative study for the argument can be done just by using nPDFs for the target and quantifying the uncertainty using a realistic nPDF set. \Figure[b]~\ref{fig:epDijet} shows the result of this exercise using EPS09LO nPDFs \cite{Eskola:2009uj} for the dijet cross section as a function of $\eta^{\mathrm{jet2}}$ where $0 < \eta^{\mathrm{jet1}} < 1$ and $x_{\gamma}^{\mathrm{obs}} > 0.75$. The latter condition reduces the contribution from resolved processes and therefore minimizes the uncertainty from the photon PDFs. With the given kinematics $\sim 10$ \% nPDF-originating uncertainty is found demonstrating the experimental accuracy required to further constrain the nPDFs. In addition, such a measurement would provide an unique test for the factorisation of the nuclear modifications. A more detailed study using the accurate photon flux from a nucleus and LHC kinematics is in the works.

\section{Summary}

We have included a framework to simulate different photoproduction processes for different collision systems for \textsc{Pythia}~8 event generator including direct and resolved processes. The framework is validated by comparing charged-particle photoproduction cross sections in e$^+$e$^-$ and ep collisions to experimental data from LEP and HERA. These data are also used to constrain the role of MPIs for the resolved photon processes for which studies have been few. The data favoured $\sim 45~(30)$ \% larger value for $p_{\mathrm{T0}}^{\mathrm{ref}}$ for $\gamma\gamma$  ($\gamma$p) than what were found optimal for pp, which translates into a smaller MPI cross-section. Also comparisons to data for photoproduction of dijets at HERA showed a reasonable agreement with the simulations. In future we will extend the photoproduction framework to ultra-peripheral heavy-ion collisions which can further constrain nPDFs.

\section*{Acknowledgements}

Work have been supported by the MCnetITN FP7 Marie Curie Initial Training Network, contract PITN-GA-2012-315877 and has received funding from the European Research Council (ERC) under the European Union's Horizon 2020 research and innovation programme (grant agreement No 668679).


\begin{thebibliography}{99}
\bibitem{Sjostrand:2014zea}
  T.~Sj\"{o}strand {\it et al.},
  Comput.\ Phys.\ Commun.\  {\bf 191} (2015) 159,
  doi:10.1016/j.cpc.2015.01.024
\bibitem{Sjostrand:2004ef}
  T.~Sj\"{o}strand and P.~Z.~Skands,
  Eur.\ Phys.\ J.\ C {\bf 39} (2005) 129,
  doi:10.1140/epjc/s2004-02084-y
\bibitem{HeleniusSjostrand}
  I.~Helenius and T.~Sj\"{o}strand,
  Work in progress.
\bibitem{Sjostrand:2004pf}
  T.~Sj\"{o}strand and P.~Z.~Skands,
  JHEP {\bf 0403} (2004) 053,
  doi:10.1088/1126-6708/2004/03/053
\bibitem{DeWitt:1978wn}
  R.~J.~DeWitt, L.~M.~Jones, J.~D.~Sullivan, D.~E.~Willen and H.~W.~Wyld, Jr.,
  Phys.\ Rev.\ D {\bf 19} (1979) 2046,
   Erratum: [Phys.\ Rev.\ D {\bf 20} (1979) 1751],
  doi:10.1103/PhysRevD.19.2046, 10.1103/PhysRevD.20.1751
\bibitem{Cornet:2002iy}
  F.~Cornet, P.~Jankowski, M.~Krawczyk and A.~Lorca,
  Phys.\ Rev.\ D {\bf 68} (2003) 014010,
  doi:10.1103/PhysRevD.68.014010
\bibitem{Ackerstaff:1998ib}
  K.~Ackerstaff {\it et al.} [OPAL Collaboration],
  Eur.\ Phys.\ J.\ C {\bf 6} (1999) 253,
  doi:10.1007/s100520050336, 10.1007/s100529801028
\bibitem{Adloff:1998vt}
  C.~Adloff {\it et al.} [H1 Collaboration],
  Eur.\ Phys.\ J.\ C {\bf 10} (1999) 363,
  doi:10.1007/s100520050761
\bibitem{Derrick:1995jq}
  M.~Derrick {\it et al.} [ZEUS Collaboration],
  Z.\ Phys.\ C {\bf 67} (1995) 227,
  doi:10.1007/BF01571283
\bibitem{Chekanov:2001bw}
  S.~Chekanov {\it et al.} [ZEUS Collaboration],
  Eur.\ Phys.\ J.\ C {\bf 23} (2002) 615,
  doi:10.1007/s100520200936
\bibitem{ATLAS:2017kwa}
  The ATLAS collaboration [ATLAS Collaboration],
  ATLAS-CONF-2017-011.
\bibitem{Eskola:2009uj}
  K.~J.~Eskola, H.~Paukkunen and C.~A.~Salgado,
  JHEP {\bf 0904} (2009) 065,
  doi:10.1088/1126-6708/2009/04/065
\end{thebibliography}
\end{document}